\documentclass[]{fairmeta}

\usepackage{iftex}
\ifPDFTeX
  \usepackage[utf8]{inputenc} 
  \usepackage[T1]{fontenc}    
\fi
\usepackage{hyperref}       
\usepackage{url}            
\usepackage{booktabs}       
\usepackage{amsfonts}       
\usepackage{nicefrac}       
\usepackage{microtype}      
\usepackage{xcolor}         
\usepackage{graphicx}
\usepackage{xspace}
\usepackage{amsmath}
\usepackage{multirow}
\usepackage{subcaption}
\usepackage{enumitem}

\newcommand{\method}{ChronoID\xspace}

\title{\method: Infusing Explicit Temporal Signals into Semantic IDs for Generative Recommendation}

\author[1]{Dongdong Nian}
\author[2]{Dongqi Fu}
\author[1]{Chenliang Xu}
\author[2]{Yinglong Xia}
\author[2]{Hong Li}
\author[2]{Hong Yan}
\author[3]{Jian Kang}

\affiliation[1]{University of Rochester}
\affiliation[2]{Meta MRS}
\affiliation[3]{MBZUAI}

\abstract{Semantic IDs are crucial in generative recommendation, but with a fundamental limitation: temporal information is not well incorporated into semantic IDs. Instead, time influences recommendation only implicitly (e.g., through session construction heuristics, preference alignment, or sequence order), while existing semantic ID learning remains entirely time-agnostic.
This design conflates interactions occurring under distinct temporal contexts into identical semantic representations, implicitly assuming that item semantics and user intent are temporally stationary. Such an assumption is misaligned with real-world recommendation scenarios, where evolving interaction rhythms play a central role.
In this work, we investigate where and how the explicit time should be incorporated into semantic ID for generative recommendation.
First, we systematically characterize the design space along three orthogonal dimensions of temporal signals and present a unified framework, \method, for time-aware semantic ID learning. Then, by contributing a new time-explicit generation recommendation benchmark, \method\ answers the questions: what is the effective way of infusing time, how to design the architecture, and where does the gain come from. 
}

\date{\today}
\correspondence{Jian Kang at \email{jian.kang@@mbzuai.ac.ae}}

\begin{document}

\maketitle

\section{Introduction}
\label{section:intro}


Recent advances in generative recommendation systems~\citep{deldjoo2024review, hou2025survey} have demonstrated the promise of replacing traditional multi-stage pipelines with end-to-end sequence generations. Methods such as OneRec~\citep{deng2025onerec} and MiniOneRec~\citep{kong2025minionerec} reformulate recommendation as a conditional generation paradigm over discrete semantic IDs (SIDs)~\citep{rajput2023tiger}, enabling the unified modeling of retrieval, ranking, and diversification. Despite their empirical success, these approaches share a critical and largely overlooked limitation: temporal information is not explicitly modeled at the level of semantic abstraction.

In existing generative recommendation frameworks, temporal signals are incorporated only implicitly or indirectly. For instance, OneRec~\citep{deng2025onerec} relies on session construction heuristics and reward modeling based on user behavior, where time influences which interactions are sampled and how preferences are aligned. However, these mechanisms operate exclusively at the data selection and optimization stages. The semantic IDs themselves, learned via vector quantization of item representations, remain fundamentally time-agnostic. As a result, temporal dynamics do not enter the tokenization layer that defines the discrete vocabulary over which generation is performed.
The open-source nature of MiniOneRec~\citep{kong2025minionerec} makes the issue even more transparent. It constructs semantic IDs by applying residual vector quantization to item textual embeddings, followed by large language model–based sequence generation.

Within this framework, temporal awareness is restricted to sequence order and relative positional encodings. While such signals capture when interactions occur within a user history, they fail to encode temporal information, i.e., the evolution of item meaning, relevance, or user intent over time. Consequently, interactions with the same item across vastly different temporal contexts are mapped to identical semantic IDs, limiting the expressiveness and adaptability of generative recommendation models.
These observations reveal a fundamental gap in current approaches:
\textit{time is modeled at the sequence and optimization levels, but excluded from semantic abstraction itself}.

Guided by this motivation, we investigate multiple architectural strategies for incorporating temporal information into semantic ID learning. These strategies reflect different inductive biases regarding the relationship between content and time, ranging from (1) time embedding schemes to (2) temporal-semantic fusion strategies and (3) codebook quantization mechanisms. By systematically exploring these designs, we aim to answer a fundamental question: \textit{Where and how should time enter semantic abstraction to enable effective generative recommendation?}
Building upon this perspective, we make the following contributions.

\textbf{First}, we present a unified framework for time-aware semantic ID learning (named \textbf{\method}) and systematically characterize its design space along three orthogonal dimensions in~\cref{sec:method}: (1) the form of temporal encoding (absolute timestamps versus relative time intervals), (2) the fusion order between temporal context and item semantics (early fusion before quantization versus late fusion at the discrete ID level), and (3) the structure of the quantizer (residual versus parallel quantization). Each design choice corresponds to a distinct inductive bias regarding how time interacts with semantic abstraction.

\textbf{Second}, to facilitate a rigorous evaluation of time-aware semantic IDs, we strictly define the time-aware generative tasks and extend the real-world time-implicit benchmarks from~\citep{hou2024bridging, kong2025minionerec} in~\cref{sec:dataset}, which not only ensures that discrete semantic IDs are learned without future information leakage during training, supervised fine-tuning, validation, and testing sets, and also provides a more rigorous and realistic evaluation for the next-item generation task.

\textbf{Third}, with the multi-design-space framework and the extended time-explicit benchmark, we conduct extensive evaluations and demonstrate that (1) incorporating time-explicit information into the semantic IDs substantially improves generative recommendation performance in~\cref{section:RQ1}; (2) relative time intervals serve as a more effective representation for modeling temporal semantic IDs compared to absolute timestamps in~\cref{section:RQ2}; (3) late fusion of time is superior to early fusion in aligning the highly heterogeneous feature spaces of item semantics and temporal signals, while more complex fusion architectures provide marginal benefits over simple concatenation in~\cref{section:RQ3}; (4) parallel quantization yields more discriminative and high-quality identifiers than residual quantization for time-aware semantic ID generation in~\cref{section:RQ4}; (5) the performance improvements are driven by enriched temporal-textual semantics rather than simply expanding the ID representation space in~\cref{appendix:Performance Improvement Resource.}; (6) temporal signals are indispensable, as removing or zero-padding them leads to significant degradation due to the introduction of out-of-distribution noise in~\cref{appendix:Necessity of Temporal Signals.}; and (7) atomic timestamps provide sufficient information for the model to internalize high-level temporal patterns (e.g., holidays and seasons) in~\cref{appendix: Impact of High-level Temporal Semantics}.

\section{Preliminary}
\label{sec:preliminary}
In this section, we provide a formal definition of the generative recommendation task and review the standard Semantic ID learning paradigm, which serves as the foundation for our proposed time-aware techniques.

\noindent \textbf{Generative Recommendation.} Let $\mathcal{U}$ and $\mathcal{I}$ denote the sets of users and items, respectively. For each user $u \in \mathcal{U}$, the interaction history is represented as a chronologically ordered sequence $S_u = (i_1, i_2, \dots, i_t)$, where $i_t \in \mathcal{I}$ is the item interacted with at timestamp $t$. The objective is to learn a model $f_\theta$ that predicts the next item $i_{n+1}$ by the conditional probability:
\begin{equation}
    P(\mathbf{z}_{n+1} \mid \mathbf{z}_{\le n}) = \prod_{k=1}^K P(z_{n+1,k} \mid \mathbf{z}_{\le n}, z_{n+1,<k}).
\end{equation}
where $\mathbf{z}_i = (z_{i,1}, z_{i,2}, \dots, z_{i,K})$ is the discrete semantic ID of item $i$, consisting of $K$ hierarchical or parallel tokens.

\noindent \textbf{Semantic ID via Quantization.} In general, semantic IDs are assigned by discretizing continuous item representations into a sequence of codebook indices. Given an item $i$ and its textual description (e.g., title, category), a pre-trained LLM is typically used to extract a high-dimensional embedding $\mathbf{e}_i \in \mathbb{R}^d$. 

A quantizer $\mathcal{Q}$ (most commonly a Residual-Quantized VAE, or RQ-VAE)~\citep{lee2022autoregressive} then maps an embeddin $\mathbf{h}_i$ to a discrete space using a set of $K$ codebooks $\{\mathcal{C}_1, \dots, \mathcal{C}_K\}$. In the residual quantization paradigm, the ID tokens are generated such that each subsequent token refines the quantization error of the previous layers:
\begin{equation}\label{eq:rq}
    \hat{\mathbf{h}}_i = \sum_{k=1}^K \mathbf{c}_{k, z_{i,k}}, 
    \quad 
    z_{i,k} = \arg\min_{j \in \{1,\ldots,V\}} \| \mathbf{r}_{k-1} - \mathbf{c}_{k,j} \|_2^2
\end{equation}
where $\mathbf{r}_{k-1}$ is the residual from the $(k-1)$-th layer and $\mathbf{c}_{k,j}$ is the $j$-th code vector in the $k$-th codebook.

However, despite its effectiveness, the aforementioned paradigm relies on a static mapping $\mathbf{h}_i \to \mathbf{z}_i$, where the embedding $\mathbf{e}_i$ is derived solely from textual content. This assumes the identity of an item in the latent space is invariant to the temporal context of the interaction. In the following sections, we describe how to move beyond this static assumption by incorporating temporal information directly into the semantic ID learning process.

\section{\method: Learning Time-Aware Semantic IDs for Recommendation}
\label{sec:method}

\begin{figure*}[h]
    \centering
    \includegraphics[width=0.86\textwidth]{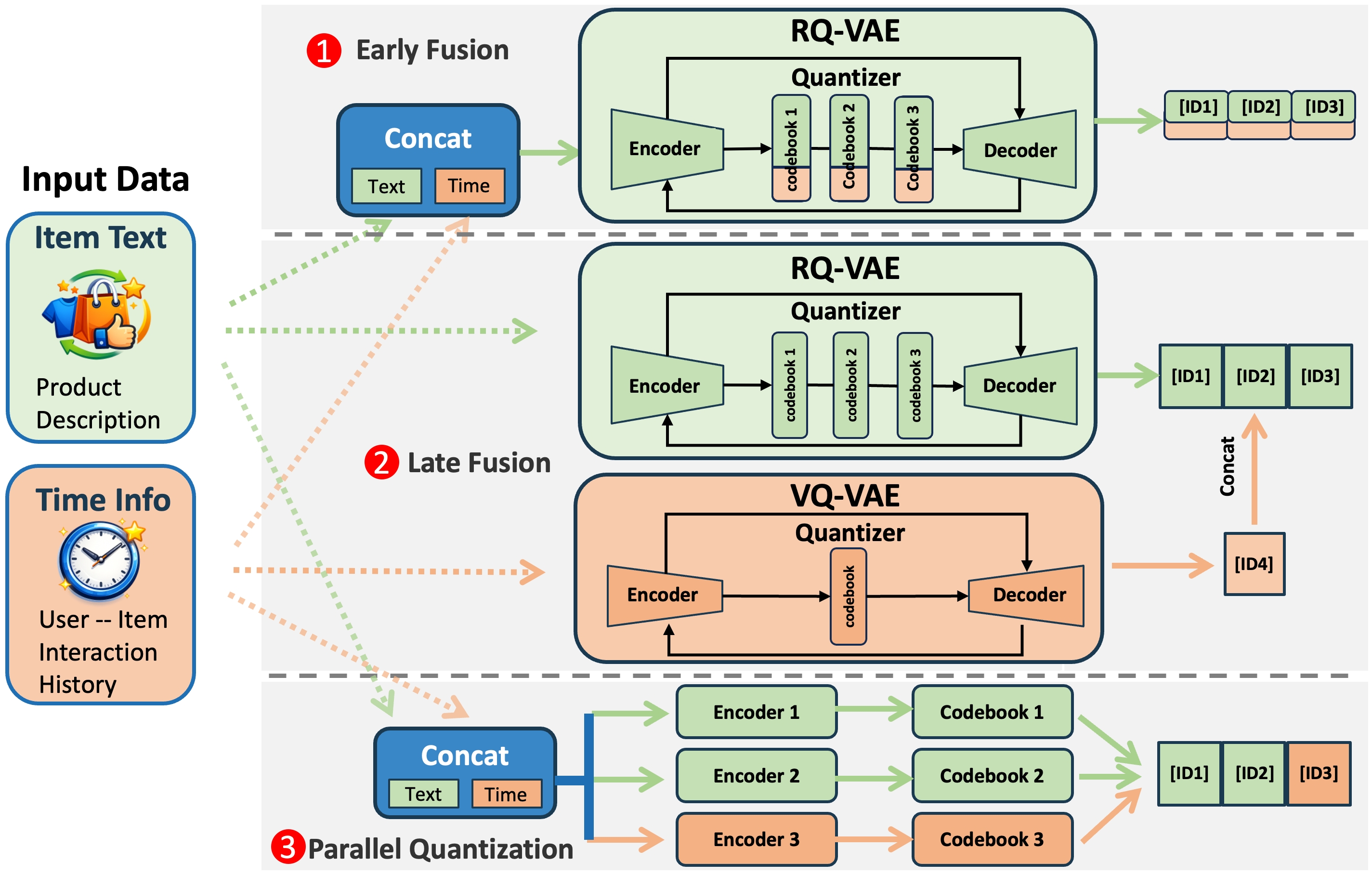}
    \caption{Three architectural variants of \method: (a) \textbf{Early Fusion}, which fuses modalities before quantization; (b) \textbf{Late Fusion}, which quantizes text and time independently; and (c) \textbf{Parallel Quantization}, which utilizes independent codebooks to capture decoupled item facets.}
    \label{fig:framework}
\end{figure*}

Here, we introduce \method, a unified framework for learning the effective time-aware semantic IDs for generative recommendation. In general, as shown in~\cref{fig:framework}, \method first produces a time embedding for each user-item interaction, then fuses the time embedding with the item embedding, and quantizes the fused embedding to extract the semantic ID via codebook training; Systematically, \method suggests that the design space of semantic ID learning involves \textbf{three key design dimensions} regarding (1) \textit{how to learn time embedding}, (2) \textit{how to fuse item embedding and time embedding}, and (3) \textit{how to quantize an embedding to obtain the semantic ID}. Finally, the semantic IDs are the inputs to the large language model for supervised fine-tuning (SFT) and inference~\citep{kong2025minionerec}. 

\subsection{Design Dimension I: Time Embedding}
Different from existing generative recommendation models that implicitly encode temporal information through sequence order, we treat temporal information as a first-class member in learning semantic IDs. Our intuition is that item semantics, such as price, seasonal trends, and user preferences could evolve over time. Thus, a high-quality encoding of temporal information could affect the quality of next-item generation, leading to a better profiling of both users and items. 

To capture the temporal relationship between events, time is first transformed into a time embedding by a time encoder. We consider the classic sinusoidal positional encoding to encode time as a temporal position of when an interaction happens~\citep{vaswani2017attention, xu2019self}. Mathematically, the $d$-dimensional time embedding $\mathbf{h}_t$ of a timestamp $t$ is computed as
\begin{equation}\label{eq:time-encoding}
\mathbf{h}_t\left[2i\right] = \sin\left(\frac{t}{10000^{2i / d}}\right),
\quad
\mathbf{h}_t\left[2i + 1\right] = \cos\left(\frac{t}{10000^{2i / d}}\right)  
\end{equation}
where $i\in\left\{1, \ldots, d\right\}$.

\noindent\textbf{Choice 1: Absolute timestamp as input.} A natural choice of input to the time encoder is the UNIX timestamp of a user interacts with an item, which we refer to as absolute time. Such a timestamp reflects the global temporal position of the event in the user-item interaction data, which captures the global trends. Besides, a key intuition of encoding time with sinusoidal function is that the inner product of two time embeddings naturally reflects the time span of two events, which helps model the frequency and recency of user-item interactions. Thus, absolute time could serve as the canonical choice of input to the time encoder.

\noindent\textbf{Choice 2: Relative time as input.} In contrast, we argue that time embedding based on absolute time may not be the optimal choice for semantic ID. The main reason is that the time embedding is later quantized into discrete semantic IDs, and the quantized semantic ID will be input to a generative model for sequence generation. During this process, there is no explicit inner product between two sinusoidal time embeddings, thus unable to correctly model the frequency and recency of user interactions. To this end, an alternative choice of input is the relative time between two consecutive interactions with respect to a user, in order to explicit encode time spans between interactions into semantic IDs. Specifically, for the $i$-th interaction made by an user $u$, we define the relative time as $\Delta t_{u, i} = t_{u, i} - t_{u, i - 1}$, 
where $t_{u, i}$ is the absolute time (i.e., UNIX timestamp) of the $i$-th interaction made by the user $u$, and $\Delta t_{u, 1} = t_{u, 1}$ is defined as the initial relative timestamp made by the user $u$. 

An illustrative example of absolute time and relative time can be found in~\cref{sec: abs_rel_time}.

\subsection{Design Dimension II: Fusion Strategy}
After temporal information is encoded, a critical design question is \textit{when} such information should be fused with item embeddings for semantic ID quantization. We distinguish between two fusion orders: (1) fuse-then-quantize and (2) quantize-then-fuse. 

\noindent \textbf{Strategy 1: Early fusion (Fuse-then-quantize).} In the early fusion setting, temporal information is treated as an inherent component of the item interaction and is fused with textual semantics prior to discretization. Suppose the item embedding of an item is $\mathbf{h}_{\text{item}}$ and the time embedding of an user interacting with this item is $\mathbf{h}_{t}$. We concatenate the time and item embedding into one embedding $\mathbf{h} = \left[\mathbf{h}_{\text{item}} || \mathbf{h}_{t}\right]$. Then we feed $\mathbf{h}$ into a vector quantization model $\mathcal{Q}$ to obtain the semantic ID $\text{SID} = \mathcal{Q}\left(\mathbf{h}\right)$.

\noindent \textbf{Strategy 2: Late fusion (Quantize-then-fuse).} In contrast to early fusion, late fusion will discretize item embedding and time embedding independently and combine them at the semantic ID level. For an item with item embedding $\mathbf{h}_{\text{item}}$ and the time embedding $\mathbf{h}_{t}$ of an user interacting with this item, we use two different vector quantization models $\mathcal{Q}_{\text{item}}$ and $\mathcal{Q}_{\text{time}}$ to obtain the quantized item ID $\text{ID}_{\text{item}} = \mathcal{Q}_{\text{item}}\left(\mathbf{h}_{\text{item}}\right)$ and time ID $\text{ID}_{\text{time}} = \mathcal{Q}_{\text{time}}\left(\mathbf{h}_{t}\right)$. Then the semantic ID $\text{SID}$ is defined as the concatenation of the item ID and time ID, i.e., $\text{SID} = \left[\text{ID}_{\text{item}} || \text{ID}_{\text{time}}\right]$.

\subsection{Design Dimension III: Quantization Mechanism}
In orthogonal to the fusion order, another key design is the quantization model. We investigate two different types of quantization models based on residual codebooks and parallel codebooks.

\noindent \textbf{Type 1: Residual quantization.} The key idea of residual quantization, such as RQ-VAE~\citep{lee2022autoregressive}, is to map an embedding to a semantic ID using $K$ different codebooks $\left\{\mathcal{C}_1, \ldots, \mathcal{C}_K\right\}$. Each codebook will be learned in a sequential order by using the residual to from the previous layer, as shown in Equation~\eqref{eq:rq}. In the case that $K = 1$ for RQ-VAE, it naturally degenerates to VQ-VAE~\citep{van2017neural}. 

\noindent \textbf{Type 2: Parallel quantization.} Different from residual quantization, parallel quantization~\citep{qu2025tokenrec} uses $K$ independent codebooks and encoders to discretize the fused representation, such that different encoders and codebooks capture different views of the input item. For a codebook in the codebook set $\mathcal{C}_k \in \{\mathcal{C}_i\}_{i = 1}^K$, it contains $V$ learnable code vectors $\mathcal{C}_k = \{\mathbf{c}_{k, 1}, \dots, \mathbf{c}_{k, V}\}$. Each codebook $\mathcal{C}_k$ independently quantizes an input embedding $\mathbf{h}$ by 
\begin{equation}
    z_{k} = \arg\min_{j \in \{1, \ldots, V\}} \| \mathbf{h} - \mathbf{c}_{k,j} \|_2^2
\end{equation}
Then the final semantic ID $\text{SID}$ of an item is the concatenation of all $z_k,~k\in\left\{1, \ldots, K\right\}$
\begin{equation}
\text{SID} = (z_1, \dots, z_K)
\end{equation}
In \method, the input to parallel quantization model is the early fused embedding $\mathbf{h} = \left[\mathbf{h}_{\text{item}} || \mathbf{h}_{t}\right]$.

\section{Time-Aware Generative Recommendation Benchmarks}
\label{sec:dataset}

To the best of our knowledge, the large-scale, text-rich, time-rigorous, and open-source recommendation datasets are limited, like Mercari~\citep{DBLP:conf/kdd/LiD0LCD25}, which is a customer-to-customer dataset with short-term temporal information.
To facilitate a rigorous evaluation, beyond Mercari~\citep{DBLP:conf/kdd/LiD0LCD25}, based on the time-implicit business-to-customer datasets (Amazon Industrial and Office \citep{hou2024bridging, kong2025minionerec}), we construct a time-explicit benchmark by splitting the original datasets using time information. Therefore, the learning of item semantic IDs is strictly partitioned by time, which ensures that discrete semantic IDs are learned without future information leakage during codebook training, supervised fine-tuning, validation, and testing, and provides a more rigorous and realistic evaluation for next-item generation.
We define a \textit{fixed} and \textit{global} temporal cutoff (e.g., 01/01/2028) to strictly partition interactions for all users. This cutoff is applied across three distinct stages to prevent any look-ahead bias:
\begin{itemize}[leftmargin=*]
\item \textbf{Codebook Training}: We strictly use interactions that occurred before the cutoff. No interactions on or after 01/01/2018, are involved in learning the item representations.
\item \textbf{Supervised Fine-Tuning (SFT) Training}: For every training instance (historical sequence, target item), we guarantee that the interaction timestamp of target item is strictly before the cutoff date.
\item \textbf{SFT Testing}: All target items to be predicted are guaranteed to occur on or after the fixed cutoff.
\end{itemize}

Under this protocol, the inclusion of users depends on their interaction timestamps relative to the cutoff. For instance, a user whose interactions all occurred before the cutoff will be used exclusively for training, whereas a user whose first interaction occurred on or after the cutoff will appear only in the test set.
Details for the \text{raw data characteristics}, \text{the formulation of training and fine-tuning datasets}, the resulting \text{data statistics}, and \text{future improvements} can be found in~\cref{appendix:dataset}.

\section{Experiments}
\label{sec:experiment}

Here, we evaluate \method to answer the following questions:
\begin{itemize}[leftmargin=*]
    \item \textbf{RQ1 (Effectiveness):} Does time-explicit semantic ID improve the performance of generative recommendation more than time-implicit baselines?
    \item \textbf{RQ2 (Time Embedding):} Comparing absolute timestamp vs. relative time, which is the better input for modeling interaction dynamics in time embedding?
    \item \textbf{RQ3 (Fusion Strategy):} Comparing early fusion vs. late fusion, which is better aligned with the hetergeneous modalities of item semantics and time?
    \item \textbf{RQ4 (Quantization Mechanism):} Is the parallel quantization better than the residual quantization for multimodal semantic ID generation?
\end{itemize}

Furthermore, we investigate whether \method's gains are driven by enriched temporal-textual semantics or simply increasing ID digits in~\cref{appendix:Performance Improvement Resource.}; we evaluate the necessity of temporal signals by comparing \method against variants where these signals are entirely removed or replaced with constant zero-padding in~\cref{appendix:Necessity of Temporal Signals.}; we explore whether explicit high-level temporal features, such as seasons and holidays, provide additional gains over raw atomic timestamps in~\cref{appendix: Impact of High-level Temporal Semantics}.

Detailed experimental configurations and hyperparameter analysis are in~\cref{appendix: Hyperparameter}.

\subsection{Experiment Setup}
To systematically answer the research questions, we decompose our framework into three key architectural dimensions. This allows us to isolate the impact of temporal encoding, fusion strategy, and quantization structure.

\paragraph{Baselines.}
We compare our framework against five representative baselines: (1) SASRec~\citep{kang2018self}, a classic self-attentive sequential model that captures user preferences from interaction sequences via unidirectional self-attention; (2) ActionPiece~\citep{hou2025actionpiece}, a recent generative recommendation method that applies BPE-style tokenization to action sequences, merging frequently co-occurring items into composite tokens; (3) HSTU~\citep{zhai2024actions}, a highly optimized transformer architecture to capture long-term user interests efficiently; (4) MiniOneRec~\citep{kong2025minionerec}, a competitive and efficient baseline for generative recommendation; and (5) TokenRec~\citep{qu2025tokenrec}, a generative recommendation framework that utilizes parallel quantizers.

\paragraph{Metrics.}
We employ two widely used top-$K$ metrics: Hit Rate (HR@K) and Normalized Discounted Cumulative Gain (NDCG@K). Following the standard leave-one-out evaluation protocol, each test instance contains exactly one ground-truth item. \textbf{HR@K} measures the proportion of test instances where the ground-truth item is present within the top-$K$ recommended list. It is defined as:
$\text{HR}@K = \frac{1}{N} \sum_{u=1}^{N} \mathbb{I}(\text{rank}_{u} \le K)$
where $N$ is the total number of users in the test set, $\text{rank}_{u}$ denotes the predicted rank of the target item for user $u$, and $\mathbb{I}(\cdot)$ is the indicator function.
\textbf{NDCG@K} considers the specific position of the ground-truth item. It assigns higher weights to items ranked closer to the top using a logarithmic decay:
$ NDCG@K = \frac{1}{N} \sum_{u=1}^{N} \frac{\log(2)}{\log(\text{rank}_{u} + 1)}$, $\text{if } \text{rank}_{u} \le K \text{ else } 0 $.
In our leave-one-out setting, since there is only one relevant item per user, the Ideal Discounted Cumulative Gain (IDCG) simplifies to $1/\log_2(1+1) = 1$, thus the denominator is omitted.


\paragraph{Time Embedding.} 
We compare two methods for encoding temporal signals into the embedding space:
\textit{Absolute Time ($T_{abs}$)}: Projects the Unix timestamp into a high-dimensional vector to capture global seasonality.
\textit{Relative Time ($T_{rel}$)}: Encodes the relative time interval between consecutive interactions, capturing the localized user rhythm.

\paragraph{Fusion Strategy.} 
This dimension investigates the optimal stage to integrate multimodal signals:
\textit{Early Fusion}: Text and time embeddings are concatenated into a single composite feature vector $\mathbf{z}_{joint} = [\mathbf{h}_{\text{text}} || \mathbf{h}_{\text{time}}]$ before being passed to the quantization layer.
\textit{Late Fusion}: Text and time are processed by independent quantization pipelines. Their respective discrete IDs are generated separately and concatenated to form the final Semantic ID.

\paragraph{Quantization Mechanism.} 
We evaluate two different discrete bottleneck structures:
\textit{Residual Codebooks}: Follows a coarse-to-fine hierarchical quantization where each subsequent codebook predicts the residual error of the previous one.
\textit{Parallel Codebooks}: Features are quantized by multiple independent codebooks in parallel, allowing different codebooks to capture decoupled item facets.

\subsection{Overall Effectiveness (RQ1)}
\label{section:RQ1}
Our main results of the effectiveness of \method is shown in~\cref{tab:main_comparison}. From the table, it is evident that \method consistently outperforms both the discriminative model (SASRec \& HSTU) and the generative recommendation baselines (ActionPiece, MiniOneRec \& TokenRec). Notably, compared to MiniOneRec, \method with parallel quantization and relative time achieves a $36.1\%$ relative improvement in HR@3 ($12.60$ vs. $9.26$) on the Industrial dataset and $40.1\%$ relative improvement in HR@3 ($8.42\%$ in HR@3 vs. $6.01\%$) on the Office dataset. These results highlight the insufficiency of solely relying on the time-implicit textual semantics and the importance of integrating explicit time in semantic IDs.

\begin{table*}[h]
    \centering
    \footnotesize 
    \setlength{\tabcolsep}{2pt} 
    \makebox[\textwidth][c]{
    \begin{tabular}{llllcccccc}
        \toprule
        \multirow{2}{*}{\textbf{Dataset}} & & \multirow{2}{*}{\textbf{Architecture}} & \multirow{2}{*}{\textbf{Strategy (Fusion, Time)}} & \multicolumn{2}{c}{\textbf{@3}} & \multicolumn{2}{c}{\textbf{@5}} & \multicolumn{2}{c}{\textbf{@10}} \\
        \cmidrule(lr){5-6} \cmidrule(lr){7-8} \cmidrule(lr){9-10}
         & & & & \textbf{HR}$\uparrow$ & \textbf{NDCG}$\uparrow$  & \textbf{HR}$\uparrow$  & \textbf{NDCG}$\uparrow$  & \textbf{HR}$\uparrow$  & \textbf{NDCG}$\uparrow$  \\
        \midrule
        
        \multirow{13}{*}{\textbf{Industrial}}
        & \multirow{5}{*}{\textbf{Baselines}} 
        & \multicolumn{1}{l}{SASRec~\citep{kang2018self}} 
        & ID-based & 7.99 & 7.12 & 9.30 & 7.68 & 10.68 & 8.46 \\
        & & \multicolumn{1}{l}{ActionPiece~\citep{hou2025actionpiece}} 
        & ID-based & 8.55 & 7.53 & 10.21 & 8.52 & 12.31 & 9.20 \\
        & & \multicolumn{1}{l}{HSTU~\citep{zhai2024actions}} 
        & Random Absolute Time & 5.11 & 3.93 & 7.05 & 4.73 & 8.64 & 5.25 \\
        & & \multicolumn{1}{l}{MiniOneRec~\citep{kong2025minionerec}} 
        & Text + Random Absolute Time & 9.26 & 8.44 & 10.95 & 9.14 & 13.53 & 10.01 \\
        & & \multicolumn{1}{l}{TokenRec~\citep{qu2025tokenrec}} & Text + Random Absolute Time & 8.20 & 7.44 & 9.01 & 7.77 & 10.54 & 8.26 \\
        \cmidrule(lr){2-10}
        
        & \multirow{6}{*}{\textbf{\method}}
        & \multirow{4}{*}{\shortstack[l]{Residual\\ Quantization}} 
          & Early, Absolute Time & 7.44 & 6.72 & 8.91 & 7.32 & 11.33 & 8.08 \\
        & \multirow{6}{*}{(Ours)} & & Early, Relative Time & 10.62 & 9.56 & 11.95 & 10.11 & 14.29 & 10.86 \\
        & & & Late, Absolute Time & 10.10 & 8.93 & 11.52 & 9.52 & 13.06 & 10.03 \\
        & & & Late, Relative Time & 10.43 & 9.28 & 11.68 & 9.79 & 13.20 & 10.29 \\
        \cmidrule(lr){3-10}
        
        & & \multirow{2}{*}{\shortstack[l]{Parallel\\ Quantization}} 
          & N/A, Absolute Time & \underline{11.22} & \underline{9.84} & \underline{12.74} & \underline{10.48} & \underline{14.77} & \underline{11.13} \\
        & & & N/A, Relative Time & \textbf{12.60} & \textbf{11.15} & \textbf{13.75} & \textbf{11.62} & \textbf{16.22} & \textbf{12.41} \\
        
        \midrule
        \midrule
        
        \multirow{13}{*}{\textbf{Office}} 
        & \multirow{5}{*}{\textbf{Baselines}}
        & \multicolumn{1}{l}{SASRec~\citep{kang2018self}} 
        & ID-based & 6.81 & 6.07 & 7.40 & 6.20 & 9.18 & 6.95 \\
        & & \multicolumn{1}{l}{ActionPiece~\citep{hou2025actionpiece}} 
        & ID-based & 5.01 & 3.49 & 5.24 & 3.28 & 8.42 & 4.30 \\
        & & \multicolumn{1}{l}{HSTU~\citep{zhai2024actions}} 
            & Random Absolute Time & 3.43 & 2.70 & 4.64 & 3.11 & 7.04 & 3.96 \\
        & & \multicolumn{1}{l}{MiniOneRec~\citep{kong2025minionerec}} & Text + Random Absolute Time & 6.01 & 4.89 & 7.22 & 5.42 & 9.53 & 6.22 \\
        & & \multicolumn{1}{l}{TokenRec~\citep{qu2025tokenrec}} & Text + Random Absolute Time & 7.54 & 6.40 & 9.10 & 7.06 & 12.10 & 8.01 \\
        \cmidrule(lr){2-10}

        & \multirow{6}{*}{\textbf{\method}}
        & \multirow{4}{*}{\shortstack[l]{Residual\\ Quantization}} 
          & Early, Absolute Time & 4.82 & 3.90 & 5.88 & 4.34 & 7.98 & 5.02 \\
        & \multirow{6}{*}{(Ours)} & & Early, Relative Time & 8.04 & \underline{6.88} & 9.46 & \underline{7.46} & 12.24 & \underline{8.34} \\
        & & & Late, Absolute Time & 7.74 & 6.66 & 9.00 & 7.20 & 10.70 & 7.74 \\
        & & & Late, Relative Time & 8.22 & 6.86 & 9.68 & \underline{7.46} & 11.71 & 8.12 \\
        \cmidrule(lr){3-10}
        
        & & \multirow{2}{*}{\shortstack[l]{Parallel\\ Quantization}} 
          & N/A, Absolute Time & \underline{7.52} & 6.14 & \underline{9.44} & 6.92 & \underline{12.30} & 7.84 \\
        & & & N/A, Relative Time & \textbf{8.42} & \textbf{7.08} & \textbf{10.74} & \textbf{8.04} & \textbf{13.59} & \textbf{8.95} \\

        \midrule
        \midrule
        
        \multirow{13}{*}{\textbf{Mercari}} 
        & \multirow{5}{*}{\textbf{Baselines}}
        & \multicolumn{1}{l}{SASRec~\citep{kang2018self}} 
        & ID-based & 0.07 & 0.04 & 0.13 & 0.06 & 0.20 & 0.09 \\
        & & \multicolumn{1}{l}{ActionPiece~\citep{hou2025actionpiece}} 
        & ID-based & 0.13 & 0.10 & 0.13 & 0.10 & 0.20 & 0.13 \\
        & & \multicolumn{1}{l}{HSTU~\citep{zhai2024actions}} 
            & Random Absolute Time & 0.02 & 0.01 & 0.02 & 0.01 & 0.03 & 0.01 \\
        & & \multicolumn{1}{l}{MiniOneRec~\citep{kong2025minionerec}} & Text + Random Absolute Time & 1.61 & 1.08 & 2.42 & 1.43 & 2.98 & 1.82 \\
        & & \multicolumn{1}{l}{TokenRec~\citep{qu2025tokenrec}} & Text + Random Absolute Time & 1.34 & 1.04 & 1.51 & 1.16 & 1.77 & 1.26 \\
        \cmidrule(lr){2-10}

        & \multirow{6}{*}{\textbf{\method}}
        & \multirow{4}{*}{\shortstack[l]{Residual\\ Quantization}} 
          & Early, Absolute Time & 1.79 & 1.41 & 2.47 & 1.65 & 3.35 & 1.97 \\
        & \multirow{6}{*}{(Ours)} & & Early, Relative Time & 2.16 & 1.69 & 2.72 & 1.91 & 3.82 & 2.28 \\
        & & & Late, Absolute Time & 1.96 & 1.58 & 2.69 & 1.81 & 3.58 & 2.10 \\
        & & & Late, Relative Time & \underline{2.49} & \underline{1.87} & 2.95 & 2.02 & \underline{3.93} & 2.30 \\
        \cmidrule(lr){3-10}
        
        & & \multirow{2}{*}{\shortstack[l]{Parallel\\ Quantization}} 
          & N/A, Absolute Time & 2.07 & 1.65 & \underline{3.08} & \underline{2.36} & 3.26 & \underline{2.74} \\
        & & & N/A, Relative Time & \textbf{3.28} & \textbf{2.59} & \textbf{4.34} & \textbf{3.03} & \textbf{5.78} & \textbf{3.50} \\
        \bottomrule
    \end{tabular}
    }
    \caption{Performance comparison on Amazon Industrial, Amazon Office, and Mercari datasets.}
    \label{tab:main_comparison}
\end{table*}

\subsection{Impact of Time Embeddings (RQ2)}
\label{section:RQ2}
For RQ2, we investigate the efficacy of different time embeddings using either absolute timestamp and relative time in \method. As shown in~\cref{tab:main_comparison}, we can observe that time embeddings based on relative time consistently and significantly outperforms its counterpart using absolute timestamp across all architectural configurations and datasets. The most distinct improvement is when early fusion and residual quantization is applied on the \textit{Industrial} dataset, which shows 42.7\% improvement for HR@3. We attribute this improvement to two major factors. First, while absolute timestamp captures global seasonality, sequential user behavior is more heavily influenced by the interaction rhythm (e.g., the gap between browsing and purchasing) which can be modeled by relative time. Second, while absolute timestamps contain seasonality, user behavior in sequential recommendation is more heavily influenced by the interval between actions (e.g., immediate re-purchase vs. periodic replacement). Absolute timestamp inherently suffers from potential distribution shift because timestamps are monotonically increasing and non-repeating. In contrast, relative time provides a more robust and generalizable representation to model the time intervals for long-term sequence generation. Both factors reflect the importance of the time intervals for modeling interaction rhythm, which are naturally supported by relative time.

\begin{figure*}[t]
    \centering
    
    \begin{subfigure}[t]{0.325\textwidth}
        \centering
        \includegraphics[width=\textwidth]{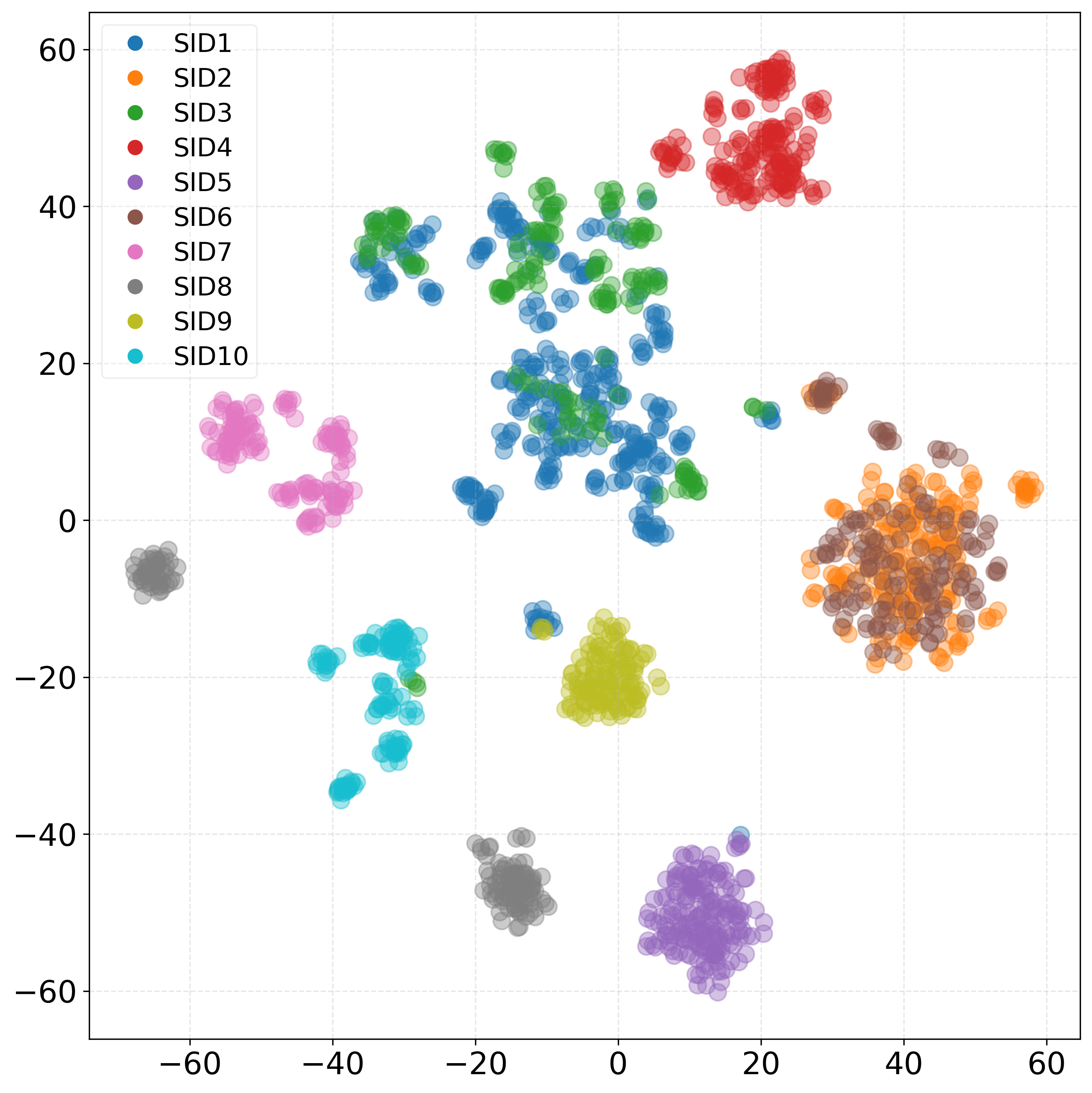}
        \caption{Parallel Quantization}
        \label{fig:tsne_mq}
    \end{subfigure}
    \hfill
    \begin{subfigure}[t]{0.325\textwidth}
        \centering
        \includegraphics[width=\textwidth]{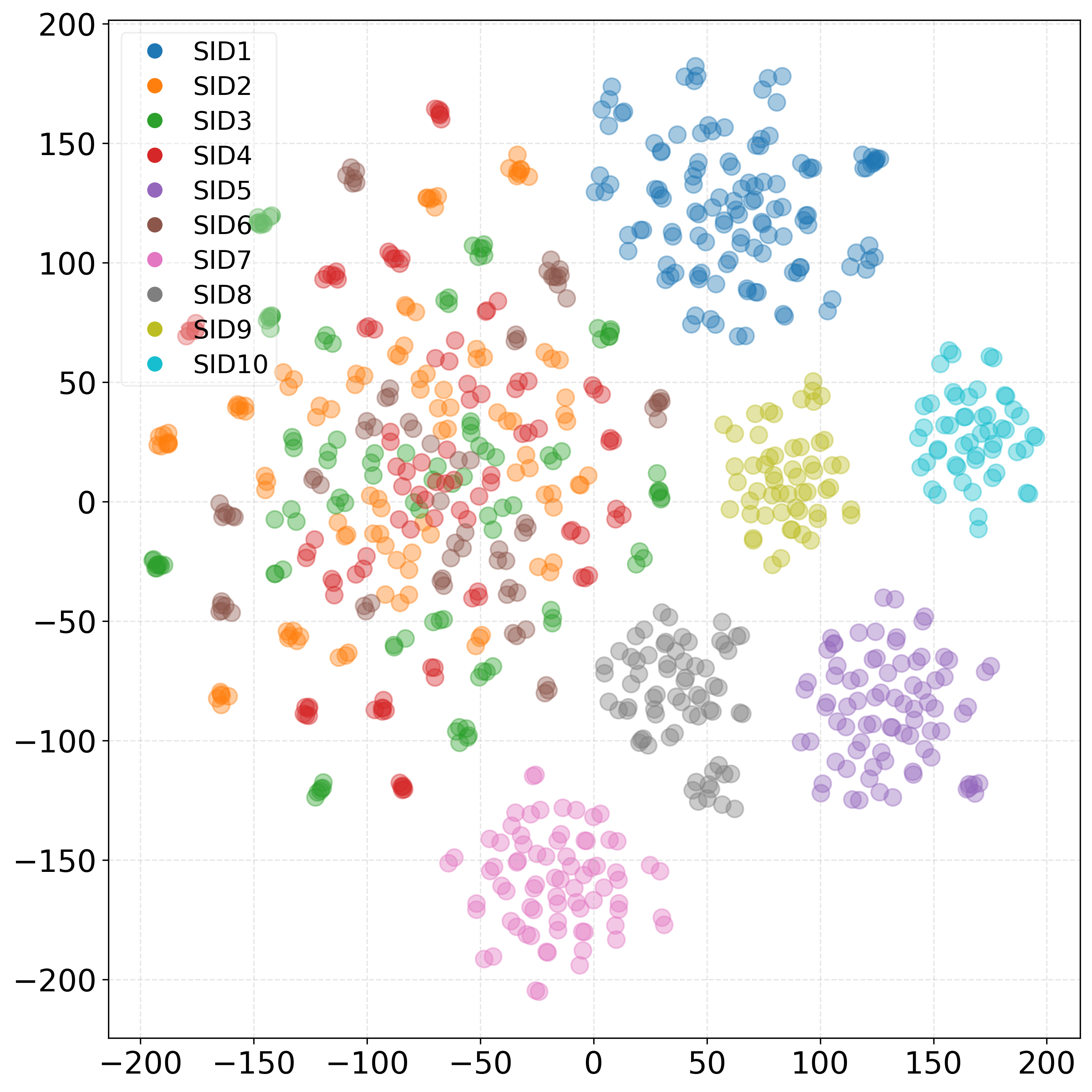}
        \caption{Residual Quantization, Absolute Time}
        \label{fig:tsne_abs}
    \end{subfigure}
    \hfill
    \begin{subfigure}[t]{0.325\textwidth}
        \centering
        \includegraphics[width=\textwidth]{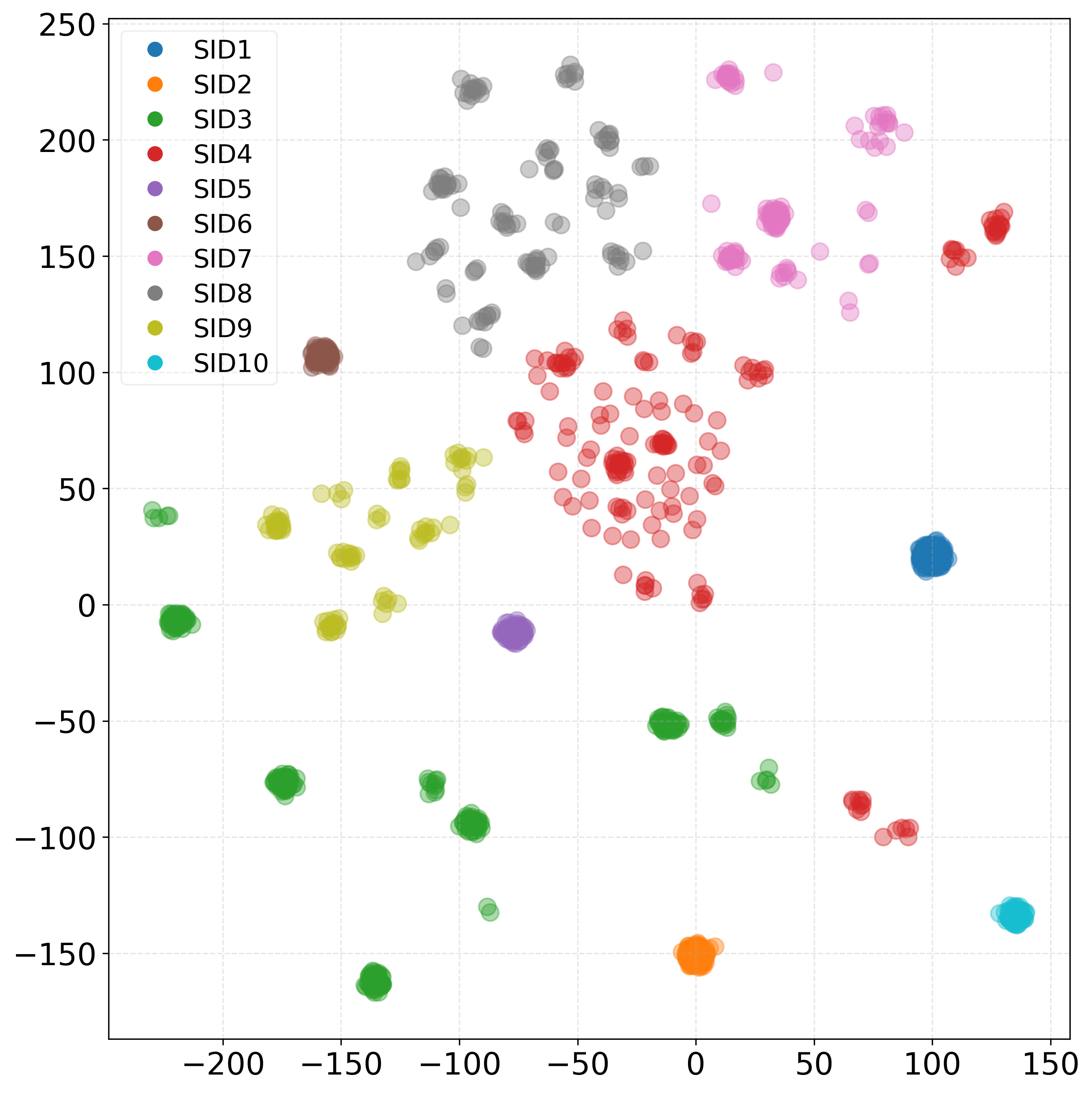}
        \caption{Residual Quantization, Relative Time}
        \label{fig:tsne_interval}
    \end{subfigure}
    
    \caption{t-SNE visualization of item embeddings for the top-10 most frequent Semantic IDs. Compared to (b) Absolute Timestamp ($T_{abs}$), the (a) Parallel Quantization strategy and the (c) Relative Time Interval ($T_{rel}$) strategy yields much tighter clusters with clearer boundaries, indicating a more discriminative semantic space.}
    \label{fig:tsne_comparison}
\end{figure*}

To further investigate how time embedings affects the learned representation space, we visualize the item embeddings associated with the top-$10$ most frequent semantic IDs using t-SNE. As illustrated in~\cref{fig:tsne_comparison} ((b) and (c)), the time embeddings generated with relative time is more clustered (i.e., clearer cluster boundaries and higher intra-cluster density) compared to those generated with absolute timestamp. In~\cref{fig:tsne_comparison} (b), clusters are more scattered and even overlap with each other more often, indicating that time embeddings from absolute timestamp is not helpful in differentiating items.

\subsection{Impact of Fusion Strategy (RQ3)}
\label{section:RQ3}
We investigate the impact of early fusion (fuse-then-quantize) and late fusion (quantize-then-fuse) for generative recommendation. From~\cref{tab:main_comparison}, we can observe that \method with late fusion consistently outperforms \method with early fusion. We attribute the substantial improvement to the fact that the textual semantics of an item and temporal information of an user-item interaction exist in highly heterogeneous feature spaces. Early fusion forces the model to compress these two distributions into a single codebook, which might lead to collapse that fails to capture the nuances of either text or time accurately. On the contrary, late fusion allow textual semantics and temporal information to preserve its unique information, ensuring that the resulting semantic ID are more informative for both modalities.

To further explore other possible early fusion methods other than the simple concatenation, we conduct additional experiments on the Industrial dataset by evaluating two alternative fusion mechanisms on relative time: (i) MLP-based, which processes concatenated text and time embeddings through a 2-layer MLP before entering the RQ-VAE encoder; and (ii) Cross-Attention, where the item text embedding acts as the Query (Q) to attend to temporal signals (K, V). Our results in~\cref{tab:fusion_ablation} reveal that these more sophisticated strategies achieve performance comparable to simple concatenation. 

\begin{table}[ht]
\centering
\resizebox{\columnwidth}{!}{
\begin{NiceTabular}{lcccccc}
\toprule
\textbf{Method} & \textbf{HR@3} & \textbf{NDCG@3} & \textbf{HR@5} & \textbf{NDCG@5} & \textbf{HR@10} & \textbf{NDCG@10} \\ \midrule
Early Fusion (Concatenation) & \textbf{10.62} & \textbf{9.56} & \textbf{11.95} & 10.11 & \textbf{14.29} & \textbf{10.86} \\
MLP-based Fusion & 10.53 & 9.47 & 11.84 & 10.16 & 13.55 & 10.62 \\
Cross-Attention Fusion & 10.72 & 9.50 & 11.05 & \textbf{10.20} & 12.49 & 10.76 \\ \bottomrule
\end{NiceTabular}
}
\caption{Ablation study on different fusion architectures on the Industrial dataset (relative time). We compare our simple concatenation with more complex fusion mechanisms.}
\label{tab:fusion_ablation}
\end{table}

\subsection{Impact of Quantization Mechanism (RQ4)}
\label{section:RQ4}
We further evaluate the impact of different quantization mechanisms, whose results are shown in Table~\cref{tab:main_comparison}. From the results, we observe that parallel quantization with relative time is consistently the best performing design choice. Apart from the advantage introduced by using relative time, we believe the fundamental advantage of parallel quantization lies in its flexibility in modeling both textual semantics and temporal information in one semantic ID. While RQ-VAE is highly effective for capturing the hierarchical, coarse-to-fine information, it enforces a rigid residual constraint where each subsequent codebook must account for the error from the previous one. However, time do not explicitly show an hierarchy, and the multimodal information (i.e., textual semantics and temporal information) are better to be characterized by independent facets rather than a hierarchy. Parallel quantization can naturally learn decoupled views of an item in different set of codebooks. This flexibility prevents error propagation in residual quantization and allows for generating more informative semantic ID for both textual semantics and temporal information when a temporal-aware generative recommendation needs to integrate heterogeneous information.

We further visualize the item embeddings generated by using residual quantization and parallel quantization with absolute timestamp using t-SNE. As shown in~\cref{fig:tsne_comparison}. Comparing~\cref{fig:tsne_comparison} (a) and~\cref{fig:tsne_comparison} (b), it is clear that parallel quantization yields more compact clusters, while clusters by residual quantization are more dispersed and intertwined. This further supports that residual quantization, which models hierarchical information, forces subsequent codebooks to model residuals that may contain noise from preceding modalities, leading to a fuzzy semantic space.

\section{Related Work}


Generative recommendation marks a paradigm shift from classical discriminative models to generating a sequence of tokens where items are represented as discrete tokens. TIGER~\citep{rajput2023tiger} is a seminal work that uses a RQ-VAE to discretize item embeddings into hierarchical semantic IDs. LETTER~\citep{wang2024letter} then improves token quality through better alignment with textual descriptions, while MiniOneRec~\citep{kong2025minionerec} investigates the scaling laws of generative recommendation and proposes a post-training pipelines for large language model-based backbones. Recent works like Forge~\citep{fu2025forge} and DAS~\citep{ye2025das} further extends the capability of discrete IDs to handle large-scale item libraries and cross-domain alignment.
A key step in generative recommendation is to discretize the embeddings via vector quantization. Majority of existing works~\citep{rajput2023tiger, kong2025minionerec} employs residual codebooks like RQ-VAE~\citep{lee2022autoregressive} to capture hierarchical semantics. TokenRec~\citep{qu2025tokenrec} instead introduces the parallel quantization strategy with multiple parallel codebooks to capture different views of an item.
Capturing the temporal information in recommendation has been a long-standing challenge. Existing works mainly employ sequence-based models for modeling temporal information. For example, GRU4Rec~\citep{hidasi2015session} uses Gated Recurrent Unit, Prompt-TPP~\citep{xue2023prompt} and ATPP~\citep{wang2021modeling} capture temporal information by modeling the temporal user-item interactions as temporal point process. 
Beyond applying the Transformer~\citep{vaswani2017attention} architecture, TiM4Rec~\citep{fan2025tim4rec} is built on the Mamba~\citep{gu2024mamba} architecture that enhances temporal modeling while preserving linear-time efficiency, it explicitly injects temporal information into the state-space formulation and achieves a better accuracy–efficiency trade-off.
TPAB~\citep{yoo2025generalizable} is architecture-agnostic and builds on standard recommendation backbones, introducing a disentangled representation framework that separates intrinsic item properties from time-varying popularity, with temporal modeling applied only to the popularity component.
Seminal work like TiSASRec~\citep{li2020time}, on the other hand, incorporate time interval information in the training data and trains a Transformer to learn the attention weights for items in the sequence, but not in a generative recommendation way.
Different from existing works in both generative recommendation and temporal recommendation, to our best knowledge, our work is the first to bridge the gap between static item semantics and dynamic nature of user-item interactions by integrating the temporal information into the semantic ID learning process.

\section{Conclusion}

In this work, we address a fundamental limitation in current generative recommendation systems: semantic IDs are learned in a time-agnostic manner. To bridge this gap, we introduce \method, a unified framework designed to learn time-aware semantic IDs. By systematically exploring a design space defined by three orthogonal dimensions, temporal encoding schemes, fusion strategies, and quantization structures, we provide a comprehensive analysis of how time should be integrated into semantic abstraction. Our empirical results on time-aware benchmarks demonstrate that incorporating temporal signals at the semantic ID level yields substantial performance gains. Specifically, our findings suggest that relative time encoding, coupled with early fusion and parallel quantization, produces the most robust and expressive semantic representations for next-item generation.



\clearpage
\newpage
\bibliographystyle{assets/plainnat}
\bibliography{paper}

\clearpage
\newpage
\beginappendix

\section{Example of Absolute Time and Relative Time}
\label{sec: abs_rel_time}
In \cref{tab:time-encoding-example}, we present an illustrative comparison between absolute time and relative time. 

\begin{table}[h]
\caption{An illustrative example of absolute timestamp ($t_{u, i}$) and relative time ($\Delta t_{u, i}$) for a sequence of three user interactions made by an user $u$. Absolute time is the same as the UNIX timestamp $t_{u, i}$, while relative time consider the time span between two consecutive interactions by the user $u$.}
\centering
\small
\begin{tabular}{c c c c}
\toprule
$i$ & UNIX Timestamp ($t_{u, i}$) & Absolute Time ($t_{u, i}$) & Relative Time ($\Delta t_{u, i}$) \\
\midrule
1 & 100 & 100 & 100 \\
2 & 115 & 115 & 15 \\
3 & 120 & 120 & 5 \\
\bottomrule
\end{tabular}

\label{tab:time-encoding-example}
\end{table}

\section{Details of Time-Aware Generative Recommendation Datasets}
\label{appendix:dataset}

\subsection{Raw Data and Feature Extraction}
We extend the Amazon Review Datasets~\citep{hou2024bridging}: \textit{Industrial} and \textit{Office}~\citep{kong2025minionerec}, which comprise user interaction logs and item metadata. For each item, we extract its textual description (e.g., title and categories) and its associated interaction timestamp.
\begin{itemize}[leftmargin=*]
\item \textbf{Textual Features:} We employ a pre-trained LLM (e.g., Qwen3-Embedding-4B~\citep{qwen3embedding-4B}) to extract a $2560$-dimensional dense embedding $\mathbf{e}_{text}$ for each item.
\item \textbf{Temporal Features:} Following the method in Section 3.1, timestamps are mapped into a $768$-dimensional vector $\mathbf{e}_{time}$ using sinusoidal encoding.
\end{itemize}

\subsection{Codebook Training: Data Formulation}
In the pre-training phase, the objective is to learn a discrete codebook that captures the joint distribution of item semantics and temporal dynamics.

\paragraph{Data Construction.}
Following our temporal isolation protocol, we filter the item library to include only those interacted with prior to 2018 ($\mathcal{I}_{past}$). As shown in~\cref{tab:dataset_stats}, this refinement yields 39,636 and 436,775 items for the \textit{Industrial} and \textit{Office} datasets, respectively. 

\paragraph{Input Representation.}
The final input to the quantizer (RQ-VAE or MQ) is a concatenated feature matrix $\mathbf{X} \in \mathbb{R}^{|\mathcal{I}_{past}| \times 3328}$, where each row $i$ corresponds to an item ID and is defined as:\begin{equation}\mathbf{x}_i = [\mathbf{e}_{text}^{(i)} \parallel \mathbf{e}_{time}^{(i)}]\end{equation}where $\parallel$ denotes the concatenation operator. This $3328$-dimensional vector serves as the ``ground truth" semantic manifold that the codebooks must reconstruct.

\subsection{Supervised Fine-tuning (SFT): Data Construction}
The SFT stage aligns the LLM's generative capacity with the learned Semantic IDs (SIDs).
\paragraph{Data Filtering.}
We implement a \textit{global temporal cutoff} (e.g., Jan 1, 2018) to strictly partition interactions across all stages. We strictly use interactions that occurred before the cutoff for codebook training. For supervised fine-tuning (SFT) training, we use training instance (historical sequence, target item), in which the interaction timestamp of target item is strictly before the cutoff.  In SFT testing, all target items to be predicted are guaranteed to occur on or after the fixed cutoff. This rigorous protocol eliminates look-ahead bias and ensures that item representations are learned without future information leakage.

\paragraph{Sample Structure.}
The SFT data is formatted as a structured dataset containing user interaction chains and their corresponding SIDs. A typical instance includes the user identity, historical titles, target title, and the discretized Semantic IDs generated by our frozen quantizer.

\subsection{Data Instances}
To provide a concrete understanding of our benchmark, we present the formatted samples for both stages.
\paragraph{Quantization Input.}
An instance in the pre-training set is a high-dimensional tensor:
\begin{quote}
\texttt{Item\_1620: [Vector(2560), Vector(768)] $\to$ Dim: 3328}
\end{quote}

\paragraph{SFT Sample Instance.}
In~\cref{tab:sft_example}, we illustrate a typical SFT record from the \textit{Industrial} dataset CSV. The model learns to map the history of Semantic IDs to the target's Semantic ID in an autoregressive manner.

\begin{table*}[t]
\centering
\resizebox{\textwidth}{!}{
\begin{tabular}{l p{4cm} p{3cm} l l}
\toprule
\textbf{User ID} & \textbf{History Titles} & \textbf{Target Title} & \textbf{History SIDs} & \textbf{Target SID} \\ 
\midrule
A5216 & [Motors 12905, Jumper Wires, ...] & yueton DPDT Toggle Switch & [\texttt{<a\_208><b\_177><c\_113>}, \dots] & \texttt{<a\_18><b\_52><c\_43>} \\ 
\bottomrule
\end{tabular}
}
\caption{Example of a Supervised Fine-tuning (SFT) data instance. The model takes a sequence of historical Semantic IDs (SIDs) as input to predict the target SID.}
\label{tab:sft_example}
\end{table*}

\subsection{Dataset Statistics}

We provide the detailed statistics of the datasets used in our experiments, comparing the original benchmark configuration with our proposed Time-aware split. As summarized in~\cref{tab:dataset_stats}, the 2018 temporal cutoff results in a more focused and challenging setup for both pre-training and supervised fine-tuning.

\begin{table}[h]
\centering
\small
\begin{tabular}{llcccc}
\toprule
\multirow{2}{*}{\textbf{Dataset}} & \multirow{2}{*}{\textbf{Split}} & \multirow{2}{*}{\textbf{Codebook-training Items}} & \multicolumn{3}{c}{\textbf{SFT Samples}} \\ \cmidrule(lr){4-6}
& & & \textbf{Train} & \textbf{Valid} & \textbf{Test} \\ \midrule
\multirow{2}{*}{\textbf{Industrial}} & Original & 43,102 & 29,446 & 3,682 & 3,683 \\
& \textbf{Time-Aware} (Ours) & 39,636 & 19,368 & 2,422 & 2,421 \\ \midrule
\multirow{2}{*}{\textbf{Office}} & Original & 472,091 & 328,422 & 41,054 & 41,054 \\
& \textbf{Time-Aware} (Ours) & 436,775 & 229,920 & 28,740 & 28,740 \\ \bottomrule
\end{tabular}%
\caption{Dataset statistics for the Original Benchmark vs. our proposed Time-aware Benchmark. The Time-aware split uses a 2018 cutoff for the Industrial dataset, significantly reducing the number of SFT samples to ensure temporal isolation and test generalization to future interactions.}
\label{tab:dataset_stats}
\end{table}

\subsection{Future Potential}

Beyond serving as a stronger static evaluation protocol, our time-aware benchmark opens several promising research directions for generative recommendation with Semantic IDs.

\paragraph{Continual Semantic ID Evolution.}
Since the quantizer is trained exclusively on historical data, the benchmark naturally supports the study of continual semantic learning. Future work can investigate how Semantic IDs should evolve when new items, modalities, or user behaviors emerge, without re-training the entire codebook from scratch. This setting enables principled comparisons between frozen, partially adaptive, and fully dynamic quantization strategies.

\paragraph{Cold-start and Emerging Item Evaluation.}
By design, all future items are unseen during the quantization phase. This makes our benchmark particularly suitable for studying cold-start and emerging item recommendation, a long-standing challenge in real-world systems. Models must infer meaningful Semantic IDs for novel items based solely on learned historical structures, closely reflecting production-level constraints.

\section{Extra Experimental Analysis}
\label{appendix:ablation study}

\subsection{Origin of Performance Gains: ID Capacity vs. Temporal Semantics}
\label{appendix:Performance Improvement Resource.}



A natural question we aim to answer is whether the performance gain of \method stems from simply expanding the ID representation space (thus reducing ID collisions) or the temporal information itself.
To investigate this question, we compare a 3-digit text-only SID with a 4-digit text-only SID on the Industrial dataset.

Ideally, the 4-digit SID significantly increases the unique representation capacity. 
However, as shown in~\cref{tab:capacity_ablation}, simply increasing the ID capacity without introducing temporal semantics leads to a performance degradation (e.g., HR@3 drops from 8.64 to 8.02). 
This result demonstrates that simply increasing the length of digits (i.e., reducing ID collisions) in a semantics-free manner is insufficient for improving recommendations.
In contrast, compared with \method, the 3-digit with time achieves substantial improvements by meaningfully integrating temporal signals, confirming that the gain is driven by the enriched temporal-textual semantics rather than mere capacity expansion.

\begin{table}[ht]
\centering
\resizebox{\columnwidth}{!}{
\begin{tabular}{lcccccc}
\toprule
\textbf{Configuration} & \textbf{HR@3} & \textbf{NDCG@3} & \textbf{HR@5} & \textbf{NDCG@5} & \textbf{HR@10} & \textbf{NDCG@10} \\ \midrule
3-digit SID (Text only) & 8.64 & 7.77 & 9.79 & 8.24 & 11.47 & 8.78 \\
4-digit SID (Text only) & 8.02 & 7.17 & 9.20 & 7.58 & 10.66 & 8.25 \\
\method (3-digit, Parallel-Rel) & \textbf{12.60} & \textbf{11.15} & \textbf{13.75} & \textbf{11.62} & \textbf{16.22} & \textbf{12.41} \\ \bottomrule
\end{tabular}
}
\caption{Impact of ID capacity and temporal semantics on the Industrial dataset. Increasing codebook capacity (from 3 to 4 digits) without temporal information fails to improve performance.}
\label{tab:capacity_ablation}
\end{table}

\subsection{Necessity of Temporal Signals: Impact of Removing and Zero-Padding} 
\label{appendix:Necessity of Temporal Signals.}
To further justify the design of \method, we conduct an experiment by comparing it with two common baselines on the Industrial dataset: (1) \textit{Remove}, which entirely omits the temporal SID digit (the standard generative recommendation approach); and (2) \textit{Replace}, which substitutes the temporal embedding with a constant zero vector while maintaining the architecture. As observed, the \textit{Remove} variant leads to a performance drop (e.g., HR@3 decreases from 10.43 to 9.47), confirming the importance of temporal context. Interestingly, the \textit{Replace} (zero-padding) strategy yields the worst performance (HR@3 of 9.04). This suggests that all-zero signals may act as out-of-distribution noise that confuses the LLM’s semantic space more than the absence of signals. These results, coupled with the fact that \method significantly outperforms the random-time baseline, further confirm that our performance gains are derived from the meaningful integration of temporal semantics rather than architectural artifacts.

\begin{table}[ht]
\centering
\resizebox{\columnwidth}{!}{
\begin{tabular}{lcccccc}
\toprule
\textbf{Variant} & \textbf{HR@3} & \textbf{NDCG@3} & \textbf{HR@5} & \textbf{NDCG@5} & \textbf{HR@10} & \textbf{NDCG@10} \\ \midrule
\method (Standard) & \textbf{10.43} & \textbf{9.28} & \textbf{11.68} & \textbf{9.79} & \textbf{13.20} & \textbf{10.29} \\
Remove (Standard Practice) & 9.47 & 8.70 & 10.70 & 9.15 & 12.58 & 9.76 \\
Replace (Zero-Padding) & 9.04 & 7.72 & 11.13 & 8.58 & 12.85 & 9.47 \\ \bottomrule
\end{tabular}
}
\caption{Sanity check on the necessity of temporal signals (Industrial dataset, relative time). We compare our standard model against variants that remove or zero-out temporal information.}
\label{tab:sanity_check}
\end{table}

\subsection{Effectiveness of Explicit High-Level Semantics} 
\label{appendix: Impact of High-level Temporal Semantics}
We further investigate whether explicitly incorporating high-level calendar-based features (e.g., weekends, seasons, and holidays) can yield additional gains. We construct a 7-digit binary indicator vector capturing: (i) weekends, (ii) the four seasons, and (iii) major holidays (e.g., Halloween, Christmas). This vector is concatenated with the original time embedding before the quantization stage. 

As shown in~\cref{tab:high_level_ablation}, the performance gains from explicit high-level tags are marginal or inconsistent across different configurations. This can be attributed to the fact that high-level semantics are deterministic functions of raw Unix timestamps. Our quantization pipeline already learns to internalize these patterns from the atomic signals, and manually adding these features may introduce redundancy, occasionally leading to slight performance degradation. 

We also observe slight improvements in specific metrics within the relative timing setting. A plausible explanation is that the discrete boundaries of temporal semantics (e.g., the start of a season) can be faded in relative representations, where adding explicit indicators may provide some compensation. However, the overall improvement remains marginal because absolute time can theoretically be recovered from relative timestamps in our SFT data (by summing the first interaction's absolute time with subsequent relative deltas). Once the absolute time is recovered, the high-level semantics are again deterministic. This confirms that our fine-grained temporal encodings are sufficient for the model to internalize high-level patterns through its underlying reasoning process.

\begin{table}[ht]
\centering
\resizebox{\textwidth}{!}{
\begin{tabular}{lllcccccc}
\toprule
\textbf{Quantization} & \textbf{Fusion / Time Type} & \textbf{Features} & \textbf{HR@3} & \textbf{NDCG@3} & \textbf{HR@5} & \textbf{NDCG@5} & \textbf{HR@10} & \textbf{NDCG@10} \\ \midrule
\multirow{4}{*}{Residual} & \multirow{2}{*}{Early, Absolute} & \method (Ours) & 7.44 & 6.72 & 8.91 & 7.32 & 11.33 & 8.08 \\
 & & + High-level Tags & 7.56 & 7.08 & 8.40 & 7.50 & 10.45 & 7.77 \\ \cmidrule(lr){2-9}
 & \multirow{2}{*}{Early, Relative} & \method (Ours) & 10.62 & 9.56 & 11.95 & 10.11 & 14.29 & 10.86 \\
 & & + High-level Tags & 10.57 & 9.78 & 11.25 & 10.06 & 11.93 & 10.28 \\ \midrule
\multirow{4}{*}{Residual} & \multirow{2}{*}{Late, Absolute} & \method (Ours) & 10.10 & 8.93 & 11.52 & 9.52 & 13.06 & 10.03 \\
 & & + High-level Tags & 10.42 & 9.48 & 11.42 & 10.25 & 12.98 & 10.17 \\ \cmidrule(lr){2-9}
 & \multirow{2}{*}{Late, Relative} & \method (Ours) & 10.43 & 9.28 & 11.68 & 9.79 & 13.20 & 10.29 \\
 & & + High-level Tags & 10.57 & 9.99 & 11.49 & 10.25 & 12.70 & 10.45 \\ \midrule
\multirow{4}{*}{Parallel} & \multirow{2}{*}{N/A, Absolute} & \method (Ours) & 11.22 & 9.84 & 12.74 & 10.48 & 14.77 & 11.13 \\
 & & + High-level Tags & 10.54 & 10.14 & 11.17 & 10.33 & 12.01 & 10.66 \\ \cmidrule(lr){2-9}
 & \multirow{2}{*}{N/A, Relative} & \method (Ours) & 12.60 & 11.15 & 13.75 & 11.62 & 16.22 & 12.41 \\
 & & + High-level Tags & \textbf{12.96} & 11.03 & \textbf{14.26} & \textbf{11.73} & \textbf{16.33} & 11.68 \\ \bottomrule
\end{tabular}
}
\caption{Ablation study on explicit high-level temporal features (Industrial dataset). "+ Tags" denotes the inclusion of explicit binary indicators for weekends, seasons, and holidays.}
\label{tab:high_level_ablation}
\end{table}

\section{Experimental Configurations and Hyperparameter Analysis}
\label{appendix: Hyperparameter}
The detailed configurations are summarized in~\cref{tab:hyperparameters} below.

\begin{table}[ht]
\centering
\small 
\setlength{\tabcolsep}{6pt} 
\begin{tabular}{llc}
\toprule
\textbf{Stage} & \textbf{Hyperparameter} & \textbf{Value} \\ \midrule
\multirow{6}{*}{Codebook Training} & Text Embedding Dimension & 2560 \\
& Temporal Encoding Dimension & 768 \\
& Number of Codebooks ($M$) & 3 \\
& Codes per Codebook ($K$) & 256 \\
& Code Dimension ($d$) & 42 \\
& Training Epochs & 10,000 \\ \midrule
\multirow{6}{*}{SFT} & Learning Rate & $3\times 10^{-4}$ \\
& Batch Size (per GPU) & 1024 \\
& Micro Batch Size & 8 \\
& Optimizer & AdamW \\
& Total Training Epochs & 10 \\ \bottomrule
\end{tabular}
\caption{Configurations for codebook training and SFT in \method.}
\label{tab:hyperparameters}
\end{table}

Here, we conduct hyperparameter analysis on two key hyperparameters in \method: the dimension of time embeddings and the number of codebooks.

\noindent \textbf{Time embedding dimension.} We investigate the sensitivity of our framework to the dimensionality of the time embeddings. This experiment is conducted on the Industrial dataset using the early fusion and relative time with 3 codebooks. As shown in~\cref{tab:dim_sensitivity}, we find that, among the tested dimensions, a dimension of $768$ offers the best results. We hypothesize that a dimension of $512$ lacks the necessary representational capacity to accurately encode fine-grained temporal information. While for the $1280$-dimensional time embedding, since the dimension of item embedding is $2560$, it is likely that an excessively large dimension for time embedding causes the model to over-focus on temporal signals while neglecting the fundamental textual semantics, which hinders the model to learn a high-quality representation that embeds both textual semantics and temporal information simultaneously.

\begin{table}[h]
\centering
\small
\begin{tabular}{lcccccc}
    \toprule
    \textbf{Dimension} & \textbf{HR@3} & \textbf{NDCG@3} & \textbf{HR@5} & \textbf{NDCG@5} & \textbf{HR@10} & \textbf{NDCG@10} \\
    \midrule
    512  & 10.08 & 9.16 & 11.60 & 9.78  & 14.15 & 10.48 \\
    \textbf{768}  & \textbf{10.62} & \textbf{9.56} & \textbf{11.95} & \textbf{10.11} & \textbf{14.29} & \textbf{10.86} \\
    1280 & 8.99  & 8.12 & 10.54 & 8.77  & 12.66 & 9.45 \\
    \bottomrule
\end{tabular}
\caption{Performance sensitivity w.r.t. time embedding dimension (Dimension) on the Industrial dataset.}
\label{tab:dim_sensitivity}
\end{table}

\noindent \textbf{Number of codebooks.} We also investigate the sensitivity of \method to the number of codebooks, which determines the length of the generated semantic IDs. We run our analysis on the \textit{Industrial} dataset using early fusion and relative time with a $768$-dimensional time embeddings. From~\cref{tab:codebook_sensitivity}, we find that the optimal performance is achieved with $3$ codebooks, while fewer codebooks or more codebooks all exhibit worse performance. 

Such a non-monotonic performance trend suggests a trade-off between representational capacity and signal-to-noise ratio in the space of semantic IDs. When there is fewer dimensions for semantic ID, the discrete encoding is insufficient to capture of complexity of the fused embeddings. The limited capacity of a 2-digit semantic ID causes significant semantic collisons, preventing the model from disinguishing between distinct user-item pairs across different times. Conversely, as the number of codebooks increases beyond the optimal point, the marginal utility of additional codebooks diminishes. For residual quantization (RQ-VAE), each subsequent layer attempts to model the remaining error from the previous layers. Thus, as we have more layers, it is more likely that the codebooks will encode noise. We empirically checked the average $\ell_1$ norm of the residual vectors in each layer and found that the norm decreases significantly as the number of codebooks increases. When the number of codebooks is $C=5$, we found that the average $\ell_1$ norm of the residual vector in the $3$rd layer is $0.36$, but the norm quickly decreases to $0.09$ in the $4$th layer and further reduces to $0.03$ in the last layer. This empirical analysis indicates that the final quantization layer, when we have more digits for the semantic IDs, is essentially attempting to discretize noise rather than meaningful semantic inforamtion, which causes a semantic saturation. Such a saturation would introduce irrelevant tokens into the semantic ID, which further degrades the overall recommendation performance.

\begin{table}[h]
\centering
\small
\begin{tabular}{lcccccc}
    \toprule
    \textbf{$C$} & \textbf{HR@3} & \textbf{NDCG@3} & \textbf{HR@5} & \textbf{NDCG@5} & \textbf{HR@10} & \textbf{NDCG@10} \\
    \midrule
    2  & 8.72  & 7.93 & 9.61  & 8.30  & 10.92 & 8.72  \\
    \textbf{3} & \textbf{10.62} & \textbf{9.56} & \textbf{11.95} & \textbf{10.11} & \textbf{14.29} & \textbf{10.86} \\
    4  & 10.54 & 9.19 & 11.67 & 9.86  & 14.15 & 10.50 \\
    5  & 9.04  & 8.39 & 10.56 & 9.01  & 12.38 & 9.60  \\
    \bottomrule
\end{tabular}
\caption{Performance sensitivity w.r.t. number of codebooks ($C$) on the Industrial Dataset.}
\label{tab:codebook_sensitivity}
\end{table}

\end{document}